\documentclass[11pt,showkeys,showpacs]{article}

\usepackage{amsmath}
\usepackage[dvips]{graphicx}

\begin{document}
\date{}
\pagestyle{myheadings}

\begin{titlepage}

\begin{flushleft}
\LARGE \textbf{On the Possible Trajectories of Particles with Spin. \\ III. Particles in the Stationary Homogeneous \\ \hspace{9mm} Electric Field}
\end{flushleft}

\parindent=40pt
\begin{minipage}[b]{120mm}

\normalsize \textbf{A.N. Tarakanov} \vskip 3mm

\small \textit{Institute of Informational Technologies, \\Belarusian State University of Informatics and Radioelectronics, \\Kozlov str. 28, 220037, Minsk, Belarus} \vskip 3mm

\small E-mail: \verb"tarak-ph@mail.ru" \vskip 10mm

\textbf{Abstract.} The behavior of spinning particles in the stationary homogeneous electric field is considered and trajectories are found for various spin orientations. We study the acceleration of spinning particles by an electric field, as well as the electric and magnetic deflection in constant homogeneous fields. It is shown that the classical Lorentz-Einstein theory of motion of charged particles is a special case of the proposed theory.
\vskip 5mm

PACS numbers: 14.60Cd, 41.90+e, 45.20.�d, 45.50.�j, 85.75.�d

Keywords: Classical mechanics, Equations of motion, Spin, Electron, Electric field

\end{minipage}
\vskip 30mm

\end{titlepage}

\section{Introduction}

Observation of the behavior of charged particles in electric and magnetic fields was made possible after enhancements by Crookes of Geissler's tube. Effect of a magnetic field on charged particles fairly easily was discovered as early as in 1821 by H.~Davy (\cite{Davy1}-\cite{Davy2}) and confirmed by J.~Pl\"{u}cker in 1858,~\cite{Plu}, W.~Hittorf in 1869, \cite{Hit}, W.~Crookes and E.~Goldstein in  1880 (\cite{Croo}, \cite{Gold}), as well as in 1878 it was discovered the action of a magnetic field on electric current in conductors (the Hall effect, \cite{HallE}). Before the discovery of the electron, physicists hardly interested in trajectories of charged particles. May be mentioned the solution of the equation of motion of charge in a homogeneous static magnetic field by E.~Ricke in 1881, \cite{Rie}, which wrote it down in a vector form with force, whose expression was later obtained by O.~Heaviside (in quaternion form, \cite{Heav}) and H. A. Lorentz (in vector form, \cite{Lor1}, p.~443; \cite{Lor2}, p.~81), and named the Lorentz force. Ricke has shown that a projection of particle trajectory onto the plane perpendicular to the field is a circle. In 1895 J. Perrin experimentally proved the~assumption of Crookes and J. J. Thomson on the corpuscular nature of cathode rays, \cite{Per1}, and showed that the negatively charged cathode particles are moving along the helical line around the direction of a~homogeneous magnetic field (\cite{Per2}, p.~497).

The action of the electric field on the cathode rays was discovered by J.~J.~Thomson only in 1897,~\cite{ThoJ23}. The failure of the earlier experiments of H.~Hertz (\cite{ThoJ23}, p.~298) for the detection of this was due to the fact that electrons of cathode rays have a very high speed, roughly measured by Thomson in 1894, which lowered pressure within the vacuum tube by means of a powerful vacuum pumps, \cite{ThoJ17}. Precise measurements of the velocity of cathode rays were performed independently by E.~Wiechert and J.~J.~Thomson in 1897 (\cite{WieE18}, \cite{ThoJ23}), which was the decisive proof that the cathode rays are electrons flow.

In 1901-1906 W. Kaufmann carried out a series of experiments on measuring of specific charge for the Becquerel $\beta$-rays (\cite{Kau1}-\cite{Kau6}). After that M. Planck has analyzed Kaufmann's data and has shown that they actually speak in favor of predictions of the Lorentz-Einstein theory rather than the theory of Abraham, who has disagreed with this conclusion (\cite{Pla1},~\cite{Pla2}). Detailed analysis of Kaufmann experiments can be found in \cite{Cush}. Henceforward most researches accept the point of view of Lorentz-Einstein. But this theory in any way does not take into account electron spin, which should be manifested also at the classical level.

An attempt to take into account spin in classical equations of motion has been undertaken in a~series author's articles, where the equations of motion, generalizing the Lorentz theory, were obtained. One of the conclusions is that the electric charge of particles should be interpreted as their helicity,~\cite{Tar1}. From this point of view, therefore, we consider that such theories of the electron with proper angular momentum as the Abraham theory, where charge is distributed in the volume of rigid sphere or ellipsoid, \cite{Abr}, or the Bucherer theory, where the electron is rotating charged sphere deformed in the oblate ellipsoid at constant volume when moving, \cite{Buch}, are not quite adequate.

This research consists of several parts, the first part of which, \cite{Tar2}, contains derivation of the~equations of motion of spinning particles in an arbitrary external field, as well as their solutions for free particles are found under the assumption that potential function depends only on the velocity of particle relative to its center of inertia. The next challenge is to find solutions for particles moving in the electric and magnetic fields. The second part, \cite{Tar3}, considers the motion in a stationary homogeneous magnetic field. This article is the third part, which aims to describe the motion of spinning particles in a stationary homogeneous electric field and compare the results with known results of the Lorentz-Einstein theory. In Sec.~2 the equations of motion of spinning particles in an external field are written in the Frenet-Serret basis, which allows considering the motion both in stationary and non-stationary fields. In Sec.~3 we study the behavior of both non-relativistic and relativistic spinless particle in a constant electric field. Sec.~4 includes solving the equations of motion and finding of trajectories of spinning particles in a constant homogeneous electric field at different orientations of spin. Assuming the longitudinal polarization of spin of free massive particle is to be conserved when the field is switched on, we consider acceleration of spinning particles in an electric field in Sec.~5, and deflection of particles in electric field corresponding to Kaufmann's experiments in Sec.~6, and deflection of particles, which are flying into the magnetic field perpendicular to the field in Sec.~7. In the concluding Sec.~8 we note some problems and make a conclusion that the use of ideas of special relativity is not mandatory when considering the motion of objects.

\section{Equations of motion of spinning particle in electric field \\ in moving reference frame}

\renewcommand{\thesection}{\arabic{section}}
\renewcommand{\theequation}{\thesection.\arabic{equation}}
\setcounter{section}{2}

We start from the equations of motion of a spinning particle in an arbitrary external field (\cite{Tar1},~\cite{Tar2})
\begin{equation}
 \frac{d}{dt} \left(m_0 \mathbf{V} - \frac{\partial U}{\partial \mathbf{V}} + \varsigma [\mathbf{s} \times \dot{\mathbf{V}}] + [\mathbf{S}^{\rm ext} \times \dot{\mathbf{V}}] \right) = \mathbf{E} + [\mathbf{V} \times (\mathbf{B} + \varsigma \Omega^2_0 \mathbf{s})] \;, \label{2.1}
\end{equation}
where
\begin{equation}
\mathbf{E} = -\frac{\partial U}{\partial \mathbf{R}} \; , \label{2.2}
\end{equation}
\begin{equation}
 U = - \int{(\mathbf{E} \cdot d\mathbf{R})} + u(t,\mathbf{V},\mathbf{\dot{V}},...,\mathbf{\dot{V}}^{(N)}) = \varphi(t,R) + u \;, \label{2.3}
\end{equation}
\begin{equation}
 \varphi(t,R) = - \int{(\mathbf{E} \cdot d\mathbf{R})} \; \label{2.4}
\end{equation}
is an electric potential of the particle at the point $\mathbf{R}$, which depends only on the relative distance $R$ in the case of constant field $\mathbf{E}$.

In what follows we assume as a first approximation that function $u$ depends only on the~velocity of particle, $u=u(V)$, what ensures conservation of total energy.

Spin components are constant in the Frenet-Serret basis (\cite{Tar1}, Appendix A). Therefore, decomposing vectors in this basis, we arrive at the equation (\cite{Tar2}, eq.~~(2.7)), which at $\mathbf{S}^{\rm ext} = \mathbf{0}$, $\mathbf{B} = \mathbf{0}$ takes the form
\begin{equation}
\begin{split}
 &\left [ \frac{d}{dt} \left (m_{0} V - \frac{du}{dV} \right ) - \varsigma s_{\rm b} (3 V\dot{V}K + V^2 \dot{K}) - E_{\rm \tau} \right ] \mathbf{e}_{\rm \tau} + \\ {} + &\left [ \left (m_{0} V - \frac{du}{dV} \right ) V K - \varsigma s_{\rm \tau} V^3 KT + \varsigma s_{\rm n} V\dot{V}T + \varsigma s_{\rm b} (\ddot{V} + \Omega^2_0 V - V^3 K^2) - E_{\rm n} \right ] \mathbf{e}_{\rm n} + \\ {} + &\left [ \varsigma s_{\rm \tau} (2 V \dot{V} K + V^2 \dot{K}) - \varsigma s_{\rm n} (\ddot{V} + \Omega^2_0 V) + \varsigma s_{\rm b} V \dot{V} T - E_{\rm b} \right ] \mathbf{e}_{\rm b} = \mathbf{0} \; . \label{2.5}
\end{split}
\end{equation}

Herewith total energy is
\begin{equation}
 \mathcal{E} = \frac{m_0 V^2}{2} - V\frac{du}{dV} + \varphi + u - \varsigma s_{\rm b} V^3 K \; . \label{2.6}
\end{equation}

Unit vectors of the Frenet-Serret basis $\mathbf{e}_{\rm \tau}$, $\mathbf{e}_{\rm n}$, $\mathbf{e}_{\rm b}$ are associated with unit vectors $\mathbf{e}_X$,~$\mathbf{e}_Y$,~$\mathbf{e}_Z$ of the absolute coordinate system by relations (\cite{Tar2}, eqs.~(2.9)-(2.11)), which, when choosing binormal direction to be fixed ($\dot{\mathbf{e}}_{\rm b} = - VT \mathbf{e}_{\rm n} = \mathbf{0}$, substituting $\Theta \rightarrow \pi/2 - \Theta$) take the form
\begin{equation}
\left \{ \begin{array}{l}
          \mathbf{e}_{\rm \tau} = \sin{\Theta(t)} \cos{\Phi} \mathbf{e}_X + \sin{\Theta(t)} \sin{\Phi} \mathbf{e}_Y + \cos{\Theta(t)} \mathbf{e}_Z \; , \\
          \mathbf{e}_{\rm n} = \cos{\Theta}(t) \cos{\Phi} \mathbf{e}_X + \cos{\Theta}(t) \cos{\Phi} \mathbf{e}_Y - \sin{\Theta}(t) \mathbf{e}_Z \; , \\
          \mathbf{e}_{\rm b} = -\sin{\Phi} \mathbf{e}_X + \cos{\Phi} \mathbf{e}_Y \; ,
        \end{array}
\right. \label{2.7}
\end{equation}
where $\dot{\Phi}=0$.  The choice $\Phi=\Phi_0=0$ corresponds to the motion that occurs in the XZ-plane, whereas $\Phi=\Phi_0=\pi/2$ corresponds to the motion in the YZ-plane. Inverse relations have the form
\begin{equation}
\left \{ \begin{array}{l}
          \mathbf{e}_X = \sin{\Theta} \cos{\Phi} \mathbf{e}_{\rm \tau} + \cos{\Theta} \cos{\Phi} \mathbf{e}_{\rm n} - \sin{\Phi} \mathbf{e}_{\rm b} \; , \\
          \mathbf{e}_Y = \sin{\Theta} \sin{\Phi} \mathbf{e}_{\rm \tau} + \cos{\Theta} \sin{\Phi} \mathbf{e}_{\rm n} + \cos{\Phi} \mathbf{e}_{\rm b} \; , \\
          \mathbf{e}_Z = \cos{\Theta} \mathbf{e}_{\rm \tau} - \sin{\Theta} \mathbf{e}_{\rm n} \; .
        \end{array}
\right. \label{2.8}
\end{equation}

Formulae (\ref{2.7}) correspond to representation of the velocity and electric force as
\begin{equation}
  \mathbf{V}(t) = V(t) \mathbf{e}_{\rm \tau} = V(t) \left[\sin{\Theta} \cos{\Phi_0} \mathbf{e}_X + \sin{\Theta} \sin{\Phi_0} \mathbf{e}_Y + \cos{\Theta} \mathbf{e}_Z \right] \; , \label{2.9}
\end{equation}
\begin{equation}
  \mathbf{E} = E_{\rm \tau} \mathbf{e}_{\rm \tau} + E_{\rm n} \mathbf{e}_{\rm n} + E_{\rm b} \mathbf{e}_{\rm b} = E_X \mathbf{e}_X + E_Y \mathbf{e}_Y + E_Z \mathbf{e}_Z \; , \label{2.10}
\end{equation}
where
\begin{equation}
\left \{ \begin{array}{l}
          E_X = E_{\rm \tau} \sin{\Theta} \cos{\Phi_0} + E_{\rm n} \cos{\Theta} \cos{\Phi_0} - E_{\rm b} \sin{\Phi_0} \; , \\
          E_Y = E_{\rm \tau} \sin{\Theta} \sin{\Phi_0} + E_{\rm n} \cos{\Theta} \sin{\Phi_0} + E_{\rm b} \cos{\Phi_0} \; , \\
          E_Z = E_{\rm \tau} \cos{\Theta} - E_{\rm n} \sin{\Theta} \; ;
        \end{array}
\right. \label{2.11}
\end{equation}
\begin{equation}
\left \{ \begin{array}{l}
          E_{\rm \tau} = E_X \sin{\Theta} \cos{\Phi_0} + E_Y \sin{\Theta} \sin{\Phi_0} + E_Z \cos{\Phi_0} \; , \\
          E_{\rm n} = E_X \cos{\Theta} \cos{\Phi_0} + E_Y \cos{\Theta} \sin{\Phi_0} - E_Z \sin{\Phi_0} \; , \\
          E_{\rm b} = -E_X \sin{\Phi_0} + E_Y \cos{\Phi_0} \; .
        \end{array}
\right. \label{2.12}
\end{equation}

Eq.~(\ref{2.9}) leads to the trajectory equation
\begin{equation}
  \mathbf{R}(t) = \mathbf{R}(0) + \int^t_0 V(t) \sin{\Theta}(t) dt (\cos{\Phi_0} \mathbf{e}_X + \sin{\Phi_0} \mathbf{e}_Y) + \int^t_0 V(t) \cos{\Theta}(t) dt \mathbf{e}_Z \; . \label{2.13}
\end{equation}

Because the torsion $T$ vanishes, and the curvature is $K=\dot{\Theta}/V$, the equation (\ref{2.5}) is simplified and looks like
\begin{equation}
\begin{split}
 &\left [ \frac{d}{dt} \left (m_{0} V - \frac{du}{dV} \right ) - \varsigma s_{\rm b} (V\ddot{\Theta} + 2 \dot{V} \dot{\Theta}) - (E_X \cos{\Phi} + E_Y \sin{\Phi}) \sin{\Theta} - E_Z \cos{\Theta} \right ] \mathbf{e}_{\rm \tau} + \\ {} + &\left [ \left (m_{0} V - \frac{du}{dV} \right ) \dot{\Theta} + \varsigma s_{\rm b} (\ddot{V} + \Omega^2_0 V - V \dot{\Theta}^2) - (E_X \cos{\Phi} + E_Y \sin{\Phi}) \cos{\Theta} + E_Z \sin{\Theta} \right ] \mathbf{e}_{\rm n} + \\ {} + &\left [ \varsigma s_{\rm \tau} (V\ddot{\Theta} + \dot{V} \dot{\Theta}) - \varsigma s_{\rm n} (\ddot{V} + \Omega^2_0 V) + E_X \sin{\Phi} - E_Y \cos{\Phi} \right ] \mathbf{e}_{\rm b} = \mathbf{0} \label{2.14}
\end{split}
\end{equation}
in the Frenet-Serret basis, or
\begin{equation}
\begin{split}
 &\left \{ \frac{d}{dt} \left [ \left (m_{0} V - \frac{du}{d V} \right ) \sin{\Theta} \right ] \cos{\Phi} - \left [ \varsigma s_{\rm \tau} (V\ddot{\Theta} + \dot{V} \dot{\Theta}) - \varsigma s_{\rm n} (\ddot{V} + \Omega^2_0 V) \right ] \sin{\Phi} \right \} \mathbf{e}_{\rm X} + \\ {} + &\left \{ \varsigma s_{\rm b} \left [ (\ddot{V} + \Omega^2_0 V - V \dot{\Theta}^2) \cos{\Theta} - (V\ddot{\Theta} + 2 \dot{V} \dot{\Theta}) \sin{\Theta} \right ] \cos{\Phi} - E_X \right \} \mathbf{e}_{\rm X} + \\ {} + &\left \{ \frac{d}{dt} \left [ \left (m_{0} V - \frac{du}{d V} \right ) \sin{\Theta} \right ] \sin{\Phi} + \left [ \varsigma s_{\rm \tau} (V\ddot{\Theta} + \dot{V} \dot{\Theta}) - \varsigma s_{\rm n} (\ddot{V} + \Omega^2_0 V) \right ] \cos{\Phi} \right \} \mathbf{e}_{\rm Y} + \\ {} + &\left \{ \varsigma s_{\rm b} \left [ (\ddot{V} + \Omega^2_0 V - V \dot{\Theta}^2) \cos{\Theta} - (V\ddot{\Theta} + 2 \dot{V} \dot{\Theta}) \sin{\Theta} \right ] \sin{\Phi} - E_Y \right \} \mathbf{e}_{\rm Y} + \\ {} + &\frac{d}{dt} \left [ \left (m_{0} V - \frac{du}{d V} \right ) \cos{\Theta} \right ] \mathbf{e}_{\rm Z} - \\ {} - &\left \{ \varsigma s_{\rm b} \left [ (\ddot{V} + \Omega^2_0 V - V \dot{\Theta}^2) \sin{\Theta} + (V\ddot{\Theta} + 2 \dot{V} \dot{\Theta}) \cos{\Theta} \right ] + E_Z \right \} \mathbf{e}_{\rm Z} = \mathbf{0} \; \label{2.15}
\end{split}
\end{equation}
in the Cartesian coordinates.

Let us choose the coordinate system so that the Z-axis to be directed along the field, i.~e.~$\mathbf{E}=E_Z \mathbf{e}_Z$, $E_X=E_Y=0$, $E_Z>0$. Then $\Theta$ is the angle between the velocity $\mathbf{V}$ and the field $\mathbf{E}$, $(\mathbf{V} \cdot \mathbf{E}) = VE_{\rm \tau} =$ $= VE_Z \cos{\Theta}$, and the electric potential is $\varphi=-\int E_Z dZ$.

Equation (\ref{2.15}) is equivalent to the system of three equations
\begin{equation}
 \frac{d}{dt} \left (m_{0} V - \frac{du}{dV} \right ) - \varsigma s_{\rm b} (V\ddot{\Theta} + 2 \dot{V} \dot{\Theta}) = E_Z \cos{\Theta} \; , \label{2.16}
\end{equation}
\begin{equation}
 \left (m_{0} V - \frac{du}{dV} \right ) \dot{\Theta} + \varsigma s_{\rm b} (\ddot{V} + \Omega^2_0 V - V \dot{\Theta}^2) = -E_Z \sin{\Theta} \; , \label{2.17}
\end{equation}
\begin{equation}
 s_{\rm \tau} (V\ddot{\Theta} + \dot{V} \dot{\Theta}) - s_{\rm n} (\ddot{V} + \Omega^2_0 V) = 0 \; \label{2.18}
\end{equation}
in three unknowns $u(V)$, $V(t)$, $\Theta(t)$. It is easy to see that eqs.~(\ref{2.16})-(\ref{2.17}) are equivalent to
\begin{equation}
 \frac{d}{dt} \left [ \left (m_{0} V - \frac{du}{dV} \right ) \cos{\Theta} - \varsigma s_{\rm b} \frac{d}{dt} (V \sin{\Theta}) - E_Z t \right ] - \varsigma s_{\rm b} \Omega^2_0 V \sin{\Theta} =0 \; , \label{2.19}
\end{equation}
\begin{equation}
 \hspace{-11mm} \frac{d}{dt} \left [ \left (m_{0} V - \frac{du}{dV} \right ) \sin{\Theta} + \varsigma s_{\rm b} \frac{d}{dt} (V \cos{\Theta}) \right ] + \varsigma s_{\rm b} \Omega^2_0 V \cos{\Theta} =0 \; . \label{2.20}
\end{equation}

Substitution of eq.~(\ref{2.17}) into eq.~(\ref{2.16}) leads to
\begin{equation}
 \varsigma s_{\rm b} \left [\frac{d}{dt} \left (\frac{\ddot{V} + \Omega^2_0 V}{\dot{\Theta}} \right ) + \dot{V} \dot{\Theta} \right ] = E_Z \left (\frac{\ddot{\Theta} \sin{\Theta}}{\dot{\Theta}^2} - 2\cos{\Theta} \right ) = -\frac{2 E_Z}{\sqrt{\dot{\Theta}}} \frac{d}{dt} \frac{\sin{\Theta}}{\sqrt{\dot{\Theta}}} \; , \label{2.21}
\end{equation}
which together with eq.~(\ref{2.18}) specifies functions $V(t)$ and $\Theta(t)$. Then eq.~(\ref{2.17}) specifies the function $u(V)$.

The equation of motion of spin
\begin{equation}
\begin{split}
 \mathbf{s} &= s_{\rm \tau} \mathbf{e}_{\rm \tau} + s_{\rm n} \mathbf{e}_{\rm n} + s_{\rm b} \mathbf{e}_{\rm b} = s_X \mathbf{e}_X + s_Y \mathbf{e}_Y + s_Z \mathbf{e}_Z = \\ {} &= (s_{\rm \tau} \sin{\Theta}\cos{\Phi_0} + s_{\rm n} \cos{\Theta}\cos{\Phi_0} - s_{\rm b} \sin{\Phi_0}) \mathbf{e}_X + \\ {} & \quad + (s_{\rm \tau} \sin{\Theta}\sin{\Phi_0} + s_{\rm n} \cos{\Theta}\sin{\Phi_0} + s_{\rm b} \cos{\Phi_0}) \mathbf{e}_Y + (s_{\rm \tau} \cos{\Theta} - s_{\rm n} \sin{\Theta}) \mathbf{e}_Z \label{2.22}
\end{split}
\end{equation}
looks like
\begin{equation}
\begin{split}
 \dot{\mathbf{s}} &= [\mathbf{\Omega}_{\rm D} \times \mathbf{s}] = \dot{\Theta} (-s_{\rm n} \mathbf{e}_{\rm \tau} + s_{\rm \tau} \mathbf{e}_{\rm n}) = \dot{\Theta} (s_{\rm \tau} \cos{\Theta} - s_{\rm n} \sin{\Theta}) \cos{\Phi_0} \mathbf{e}_X + \\ {} &\quad + \dot{\Theta} (s_{\rm \tau} \cos{\Theta} - s_{\rm n} \sin{\Theta}) \sin{\Phi_0} \mathbf{e}_Y - \dot{\Theta} (s_{\rm \tau} \sin{\Theta} + s_{\rm n} \cos{\Theta}) \mathbf{e}_Z \; , \label{2.23}
\end{split}
\end{equation}
i.~e. spin precesses with angular velocity
\begin{equation}
 \mathbf{\Omega}_{\rm D} = \dot{\Theta} \mathbf{e}_{\rm b} = \dot{\Theta} (- \sin{\Phi_0} \mathbf{e}_X + \cos{\Phi_0} \mathbf{e}_Y) \label{2.24}
\end{equation}
around the binormal direction $\mathbf{e}_{\rm b}$.

Eqs.~(\ref{2.16})-(\ref{2.18}) does not contain any hints as to what values can take spin components. Therefore, we consider below different possible cases.

\section{Spinless particle in constant electric field}

\renewcommand{\thesection}{\arabic{section}}
\renewcommand{\theequation}{\thesection.\arabic{equation}}
\numberwithin{equation}{section}
\setcounter{section}{3}
\setcounter{equation}{0}

Let us consider the classic case of absence of spin $s_{\rm \tau}=s_{\rm n}=s_{\rm b}=0$. Equation (\ref{2.1}) with taking into account of $u=u(V)$ leads to
\begin{equation}
 \left (m_{0} - \frac{du}{VdV} \right ) \mathbf{V} - \mathbf{E}t = m_0 \gamma_0 \mathbf{V}_0 = \mathbf{const} \; , \; \; \gamma_0 = \mathrm{const} \; , \label{3.1}
\end{equation}
wherefrom
\begin{equation}
 m_{0} V - \frac{du}{dV} = \pm \sqrt{m^2_0 \gamma^2_0 \mathbf{V}^2_0 + 2 m_0 \gamma_0 (\mathbf{V}_0 \cdot \mathbf{E}) t + \mathbf{E}^2 t^2} \; . \label{3.2}
\end{equation}
Eq.~(\ref{3.2}) can be solved, if we know the dependence of the function $u(V)$ on the velocity. Apart from the theories of Abraham and Bucherer, we know two cases of such dependence, corresponding to: 1)~non-relativistic motion, $u=0$, $\gamma_0=1$, and 2)~relativistic motion in the Lorentz-Einstein theory, in which we have to put
\begin{equation}
 m_{0} - \frac{du}{VdV} = m_0 \gamma_{\rm L}(V) = \frac{m_0}{\sqrt{1 - V^2 / c^2}} \; , \label{3.3}
\end{equation}
from which we find
\begin{equation}
 u(V) = \frac{m_0 V^2}{2} + m_0 c^2 \sqrt{1 - V^2 / c^2} \; . \label{3.4}
\end{equation}

\textbf{I.1.} In the non-relativistic case $u=0$ and eq.~(\ref{3.1}) becomes $\mathbf{V}(t)=\mathbf{V}_0+\mathbf{E}t/m_0$, which yields the vector equation of trajectory
\begin{equation}
 \mathbf{R}(t) = \mathbf{R}(0) + \mathbf{V}_0 t + \mathbf{E}t^2/2 m_0 \; , \label{3.5}
\end{equation}
or in parametric form
\begin{equation}
  \left \{
  \begin{array}{l}
          X(t) = X(0) + V_0 t \sin{\Theta_0}\cos{\Phi_0} \; , \\
          Y(t) = Y(0) + V_0 t \sin{\Theta_0}\sin{\Phi_0} \; ,\\
          Z(t) = Z(0) + V_0 t \cos{\Theta_0} + E_Z t^2/2 m_0 \; ,
        \end{array}
        \right. \; \label{3.6}
\end{equation}
where $\Theta_0$, $\Phi_0$ are angles at the initial time $t=0$. If we choose the coordinate axes so that the initial position of the particle has coincided with the origin, i.~e. $\mathbf{R}(0)=\mathbf{0}$, and get rid of the parameter $t$ in eq.~(\ref{3.6}), then we obtain the trajectory in the form of canonical equation of parabola
\begin{equation}
 X^2 +Y^2 + 2A\sqrt{X^2 +Y^2} = W^2 - A^2 = BZ \; , \label{3.7}
\end{equation}
where
\begin{equation}
 W = \sqrt{X^2 +Y^2} + A \; , \quad A = \frac{m_0 V^2_0 \sin{2\Theta_0}}{2E_Z} \; , \quad B = \frac{2m_0 V^2_0 \sin^2{\Theta_0}}{E_Z} \; .\label{3.8}
\end{equation}

Note that in the case of classical Lorentz force in eq.~(\ref{2.1}) we use the electric force $\mathbf{E}$ and the magnetic vector $\mathbf{B}$ instead of conventional expressions $e\mathbf{E}$ and $e\mathbf{B}$, where $e$ is electric charge, and $\mathbf{E}$ and $\mathbf{B}$ are electric field strength and magnetic field (magnetic flux density), respectively. Therefore, one should put $E_Z=eE$ in eqs.~(\ref{3.6})-(\ref{3.8}) for the classical solution. Hence the trajectory of the particle with opposite charge (i.~e., the opposite helicity) is obtained from eqs.~(\ref{3.6})-(\ref{3.8}), if $Z$ replaced by $-Z$.

\textbf{I.2.} In the relativistic case $u(V)$ is given by eq.~(\ref{3.4}), and eq.~(\ref{3.1}) becomes
\begin{equation}
 m_{0} \gamma_{\rm L} \mathbf{V} = m_{0} \gamma_{\rm L} \frac{d \mathbf{R}}{dt} = m_0 \gamma_0 \mathbf{V}_0 + \mathbf{E} t = \frac{m_0 \mathbf{V}_0}{\sqrt{1-V^2_0/c^2}} + \mathbf{E} t \; , \label{3.9}
\end{equation}
from whence we obtain for the Lorentz factor
\begin{equation}
 \gamma_{\rm L} = \sqrt{\frac{\mathbf{E}^2}{m^2_0 c^2} t^2 +\frac{2 \gamma_0 (\mathbf{V}_0 \cdot \mathbf{E})}{m_0 c^2} t + \gamma^2_0} = \frac{E_Z}{m_0 c} \sqrt{(t+t_1)^2 +t^2_2} \; , \label{3.10}
\end{equation}
where
\begin{equation}
 t_1 = \frac{m_0 \gamma_0 (\mathbf{V}_0 \cdot \mathbf{E})}{\mathbf{E}^2} \; , \quad t^2_2 = \frac{m^2_0 c^2 \gamma^2_0}{\mathbf{E}^2} \left (1 - \frac{(\mathbf{V}_0 \cdot \mathbf{E})^2}{c^2 \mathbf{E}^2} \right ) \; , \quad t^2_1 + t^2_2 = \frac{m^2_0 c^2 \gamma^2_0}{\mathbf{E}^2} \; . \label{3.11}
\end{equation}
The substituting of eq.~(\ref{3.10}) into eq.~(\ref{3.9}) and subsequent integration gives the vector equation of the trajectory
\begin{equation}
\begin{split}
 \mathbf{R}(t) &= \mathbf{R}(0) + \gamma_0 \mathbf{V}_0 \int^t_0 \frac{dt}{\gamma_{\rm L}} + \frac{\mathbf{E}}{m_0} \int^t_0 \frac{t dt}{\gamma_{\rm L}} = \mathbf{R}(0) + \frac{c \mathbf{E}}{E_Z} \left [\sqrt{(t+t_1)^2 +t^2_2} - \sqrt{t^2_1 +t^2_2} \right ] + \\ {} & \quad + \frac{m_0 c \gamma_0 [\mathbf{E} \times [ \mathbf{V}_0 \times \mathbf{E}]]}{E^3_Z} \ln{\left [1 + \frac{t}{t_1 + \sqrt{(t+t_1)^2 +t^2_2}} \right ]} \; , \label{3.12}
\end{split}
\end{equation}
or in parametric form
\begin{equation}
  \left\{
    \begin{array}{ll}
      X(t) = X(0) + \cfrac{m_0 c V_0 \gamma_0}{E_Z} \sin{\Theta_0}\cos{\Phi_0} \ln{\left [1 + \cfrac{t}{t_1 + \sqrt{(t+t_1)^2 +t^2_2}} \right ]} \; , \\
      Y(t) = Y(0) \; + \cfrac{m_0 c V_0 \gamma_0}{E_Z} \sin{\Theta_0}\sin{\Phi_0} \ln{\left [1 + \cfrac{t}{t_1 + \sqrt{(t+t_1)^2 +t^2_2}} \right ]} \; , \\
      Z(t) = Z(0) + c \left [\sqrt{(t+t_1)^2 +t^2_2} - \sqrt{t^2_1 +t^2_2} \right ] \; .
    \end{array}
  \right.
\label{3.13}
\end{equation}

\textbf{I.3.} Assuming that the function $u(V)$ is non-zero and does not determined by the relation eq.~(\ref{3.4}), and trying to define it from eqs.~(\ref{2.16})-(\ref{2.18}), it is easy to see that at zero spin eq.~(\ref{2.18}) becomes an identity, and eqs.~(\ref{2.16})-(\ref{2.17}) are equations for three unknown functions $u(V)$, $\Theta(t)$ and $V(t)$. Therefore zero spin, and hence the Lorentz equations of motion, does not give possibility to determine the dependence of the function $u(V)$ on the velocity.

Substituting eq.~(\ref{2.17}) into eq.~(\ref{2.16}) at $s_{\rm b}=0$, we get
\begin{equation}
 \frac{d}{dt} \left (m_{0} V - \frac{du}{dV} \right ) = -E_Z \cos{\Theta} + E_Z \frac{\ddot{\Theta} \sin{\Theta}}{\dot{\Theta}^2} = E_Z \cos{\Theta} \; , \label{3.14}
\end{equation}
whence
\begin{equation}
 \frac{\ddot{\Theta}}{\dot{\Theta}} = \frac{d(\ln{\dot{\Theta}})}{dt} = 2\dot{\Theta} \cot{\Theta} = \frac{d(\ln{\sin^2{\Theta}})}{dt} \; , \quad \dot{\Theta} = \lambda \sin^2{\Theta} \; , \quad \lambda = \mathrm{const} \; , \label{3.15}
\end{equation}
\begin{equation}
 \Theta(t) = \arctan^{-1}{(\cot{\Theta_0} - \lambda t)} \; , \label{3.16}
\end{equation}
and
\begin{equation}
 \sin{\Theta} = \frac{1}{\sqrt{1 + (\cot{\Theta_0} - \lambda t)^2}} \; , \quad \cos{\Theta} = \frac{\cot{\Theta_0} - \lambda t}{\sqrt{1 + (\cot{\Theta_0} - \lambda t)^2}} \; . \label{3.17}
\end{equation}

Substituting eqs.~(\ref{3.15}), (\ref{3.17}) into eq.~(\ref{2.17}) and introducing the function $\gamma(V)$ instead of $u(V)$, we arrive at equation
\begin{equation}
 m_{0} V - \frac{du}{dV} = m_{0} \gamma V = -\frac{E_Z}{\lambda} \sqrt{1 + (\cot{\Theta_0} - \lambda t)^2} \; \label{3.18}
\end{equation}
for two unknowns $V(t)$ and $u(V)$ (or $\gamma(V)$). If
\begin{equation}
 \lambda = -\frac{E_Z}{m_{0} \gamma_0 V_0 \sin{\Theta_0}} \; , \label{3.19}
\end{equation}
then eq.~(\ref{3.18}) describes both non-relativistic case ($u=0$, $\gamma=\gamma_0=\gamma_{\rm L}(0)=1$) and relativistic one (eq.~(\ref{3.4}), $\gamma=\gamma_{\rm L}(V)$). When the initial velocity is directed along ($\Theta_0=0$) or antiparallel the field $\mathbf{E}$ ($\Theta_0=\pi$), then $\lambda=-\infty$.

\section{Spinning particle in constant electric field}

\renewcommand{\thesection}{\arabic{section}}
\renewcommand{\theequation}{\thesection.\arabic{equation}}
\setcounter{section}{4}
\setcounter{equation}{0}

Let us now find the particle trajectory, if its spin components (\ref{2.22}) do not equal zero.

\textbf{II.} $s_{\rm \tau}=0$, $s_{\rm n} = 0$, $s_{\rm b} \neq 0$. Here eq.~(\ref{2.18}) becomes an identity, so that as a result we have  two equations (\ref{2.16}) and (\ref{2.17}) relative to three unknown functions $u(V)$, $\Theta(t)$ and $V(t)$. Hence, the definition of two functions is impossible, if the third function is unknown. In the limit $s_{\rm b}=0$ the solution should reduce to eq.~(\ref{3.5}), when $u=0$, or to eq.~(\ref{3.12}), if $u$ is given in eq.~(\ref{3.4}).

Eqs.~(\ref{2.16})-(\ref{2.18}), which can be conveniently represented in Cartesian coordinates, is reduced to a system of three equations
\begin{equation}
 \frac{d}{dt} \left [ \left (m_{0} - \frac{du}{VdV} \right ) V_X \right ] + \varsigma s_{\rm b} \cos{\Phi_0} (\ddot{V}_Z + \Omega^2_0 V_Z) = 0  \; , \label{4.1}
\end{equation}
\begin{equation}
 \frac{d}{dt} \left [ \left (m_{0} - \frac{du}{VdV} \right ) V_Y \right ] + \varsigma s_{\rm b} \sin{\Phi_0} (\ddot{V}_Z + \Omega^2_0 V_Z) = 0 \; , \label{4.2}
\end{equation}
\begin{equation}
 \frac{d}{dt} \left [ \left (m_{0} - \frac{du}{VdV} \right ) V_Z \right ] - \varsigma s_{\rm b} (\ddot{V}_{(XY)} + \Omega^2_0 V_{(XY)}) = E_Z \; , \label{4.3}
\end{equation}
only two of which are independent. Indeed, multiplying eqs.~(\ref{4.1}) and (\ref{4.2}) by $\sin{\Phi_0}$ and $\cos{\Phi_0}$, and then adding and subtracting the results, we will come to equations
\begin{equation}
 \frac{d}{dt} \left [ \left (m_{0} - \frac{du}{VdV} \right ) V_{[XY]} \right ] = 0  \; , \label{4.4}
\end{equation}
\begin{equation}
 \frac{d}{dt} \left [ \left (m_{0} - \frac{du}{VdV} \right ) V_{(XY)} \right ] + \varsigma s_{\rm b} (\ddot{V}_Z + \Omega^2_0 V_Z) = 0  \; , \label{4.5}
\end{equation}
where
\begin{equation}
 V_{[XY]}(t) = V_X(t) \sin{\Phi_0} - V_Y(t) \cos{\Phi_0} \; , \quad V_{(XY)}(t) = V_X(t) \cos{\Phi_0} + V_Y(t) \sin{\Phi_0} \; , \label{4.6}
\end{equation}
and eq.~(\ref{4.4}) becomes an identity, forasmuch as $V_X$, $V_Y$ and $V_Z$ are expressed in terms of $V(t)$, $\Theta(t)$ and $\Phi_0$ according to eq.~(\ref{2.9}).

By differentiating eqs.~(\ref{4.3}) and (\ref{4.5}), and multiplying them by $m_0$, and combining the resulting equations, we come to differential equations
\begin{equation}
\begin{split}
 \frac{d}{dt} &\left [ \frac{d^2}{dt^2} \left ( \frac{du}{V dV} V_Z \right ) + \Omega^2_0 \frac{du}{V dV} V_Z - \frac{m_0}{\varsigma s_{\rm b}} \frac{d}{dt} \left ( \frac{du}{V dV} V_{(XY)} \right ) \right ] + \\ {} &+ \varsigma s_{\rm b} \left [ V^{(4)}_{(XY)} + 2 \Omega^2 \ddot{V}_{(XY)} + \Omega^4_0 (V_{(XY)} - V_1) \right ] = 0  \; , \label{4.7}
\end{split}
\end{equation}
\begin{equation}
\begin{split}
 \frac{d}{dt} &\left [ \frac{d^2}{dt^2} \left ( \frac{du}{V dV} V_{(XY)} \right ) + \Omega^2_0 \frac{du}{V dV} V_{(XY)} + \frac{m_0}{\varsigma s_{\rm b}} \frac{d}{dt} \left ( \frac{du}{V dV} V_Z \right ) \right ] - \\ {} & - \varsigma s_{\rm b} \left [ V^{(4)}_Z + 2 \Omega^2 \ddot{V}_Z + \Omega^4_0 V_Z \right ] = 0 \; , \label{4.8}
\end{split}
\end{equation}
where
\begin{equation}
 \Omega^2 = \Omega^2_0  + \omega^2_0 \; , \quad \omega^2_0 = \frac{m^2_0}{2 \varsigma^2 s^2_{\rm b}} \; , \quad V_1 = - \frac{E_Z}{\varsigma s_{\rm b} \Omega^2_0} \; . \label{4.9}
\end{equation}

In general, it is very difficult to analyze the system (\ref{4.7})-(\ref{4.8}). However, it is easy to make for $u=0$. Then the general solution of eqs.~(\ref{4.7})-(\ref{4.8}) is determined by the roots of characteristic equation
\begin{equation}
 \lambda^4 + 2 \Omega^2 \lambda^2\ + \Omega^4_0 = 0 \; , \label{4.10}
\end{equation}
which are imaginary
\begin{equation}
 \lambda_{1,2} = \pm i \sqrt{\Omega^2 + \sqrt{\Omega^4 - \Omega^4_0}} = \pm i \Omega_{+} \; , \quad \lambda_{3,4} = \pm i \sqrt{\Omega^2 - \sqrt{\Omega^4 - \Omega^4_0}} = \pm i \Omega_{-} \; . \label{4.11}
\end{equation}
Hence, the general solution of eqs.~(\ref{4.7})-(\ref{4.8}) is
\begin{equation}
 V_{(XY)}(t) = V_1 + V^{+}_{XY} \cos{(\Omega_{+} t + \phi_{+})} + V^{-}_{XY} \cos{(\Omega_{-} t + \phi_{-})} \; , \quad V^{\pm}_{XY} = \mathrm{const} \; , \label{4.12}
\end{equation}
\begin{equation}
 V_Z (t) = V^{+}_Z \cos{(\Omega_{+} t + \varphi_{+})} + V^{-}_Z \cos{(\Omega_{-} t + \varphi_{-})} \; , \quad V^{\pm}_{Z} = \mathrm{const} \; . \label{4.13}
\end{equation}

Substituting eqs.~(\ref{4.12})-(\ref{4.13}) into input equations to determine correlation between the integration constants, we find that the system (\ref{4.1})-(\ref{4.3}) (at $u=0$) has a nontrivial solution only at $s_{\rm b}=0$. Consequently, there are no trajectories at $s_{\rm \tau}=0$, $s_{\rm n}=0$, $s_{\rm b}=0$, $u=0$. Apparently, this also occurs when $u \neq 0$. Proof of the existence or absence of trajectories is not currently feasible.

\vspace{3mm}
\textbf{III.} $s_{\rm \tau} \neq 0$, $s_{\rm n}=0$, $s_{\rm b}=0$. It is shown in \cite{Tar2} that the spin of free massive particle is arranged parallel or antiparallel to the velocity. Assuming this polarization is to be conserved when the field is switched, from eq.~(\ref{2.18}) we obtain $V\ddot{\Theta} + \dot{V}\dot{\Theta} = 0$, from where $\dot{\Theta} = \lambda V_0/V$, where $\lambda \neq 0$ is an integration constants. As a result, the function $\Theta(t)$ depends on time according to eq.~(\ref{3.16}). From eqs.~(\ref{2.9}), (\ref{3.15}) and (\ref{3.17}) we find
\begin{equation}
 V(t) = \frac{V_0}{\sin^2{\Theta(t)}} = V_0 [1 + (\cot{\Theta_0} - \lambda t)^2] \; , \label{4.14}
\end{equation}
\begin{equation}
 \mathbf{V}(t) = V_0 \sqrt{1 + (\cot{\Theta_0} - \lambda t)^2} \left [ \cos{\Phi_0} \mathbf{e}_X + \sin{\Phi_0} \mathbf{e}_Y + (\cot{\Theta_0} - \lambda t) \mathbf{e}_Z \right ] \; . \label{4.15}
\end{equation}

The solution (\ref{4.15}) can be represented as eq.~(\ref{3.9}), where $\lambda$ is given in eq.~(\ref{3.19}), but $\gamma(V)$ in contrast to the Lorentz factor $\gamma_{\rm L}(V)$ in eq.~(\ref{3.3}) depends on the velocity by the law
\begin{equation}
 \gamma(V) = 1 - \frac{du}{m_0 VdV} = \frac{\gamma_0 \sin{\Theta_0}}{\sqrt{V/V_0}} = - \frac{E_Z}{\lambda m_0 \sqrt{V_0 V}} \; , \label{4.16}
\end{equation}
which can be derived also from the equation for function $u(V)$, obtained by substitution of eqs.~(\ref{3.16}) and (\ref{4.14}) into eq.~(\ref{2.17})
\begin{equation}
 \frac{du}{dV} = m_0 V + \frac{E_Z}{\dot{\Theta}} \sin{\Theta} = m_0 V + \frac{E_Z}{\lambda} \sqrt{\frac{V}{V_0}} \; , \label{4.17}
\end{equation}
wherefrom
\begin{equation}
 u(V) = \frac{m_0 V^2}{2} + \frac{2E_Z V^{3/2}}{3 \lambda \sqrt{V_0}} + u(0) \; . \label{4.18}
\end{equation}

The dependence (\ref{4.16}) shows that the effective mass decreases when the velocity increases. Thus, the trajectory equation differs from eq.~(\ref{3.12}) and looks like
\begin{equation}
\begin{split}
 \mathbf{R}(t) &= \mathbf{R}(0) - \frac{V_0}{2 \lambda} \left [ (\cot{\Theta_0} - \lambda t) \sqrt{1 + (\cot{\Theta_0} - \lambda t)^2} - \frac{\cos{\Theta_0}}{\sin^2{\Theta_0}} \right ] \left ( \cos{\Phi_0} \mathbf{e}_X + \sin{\Phi_0} \mathbf{e}_Y \right ) - \\ {} & \quad - \frac{V_0}{2 \lambda} \ln{ \left \{ \frac{\sin{\Theta_0}}{1 + \cos{\Theta_0}} \left [\cot{\Theta_0} - \lambda t + \sqrt{1 + (\cot{\Theta_0} - \lambda t)^2} \; \right ] \right \} } \left (\cos{\Phi_0} \mathbf{e}_X + \sin{\Phi_0} \mathbf{e}_Y \right ) + \\ {} & \quad + \frac{V_0}{3 \lambda} \left [ \frac{1}{\sin^3{\Theta_0}} - \left [1 + (\cot{\Theta_0} - \lambda t)^2 \right ]^{3/2} \right ] \mathbf{e}_Z \; . \label{4.19}
\end{split}
\end{equation}

Spin $\mathbf{s} = s_{\rm \tau} \mathbf{e}_{\rm \tau}$, being always directed along the tangent to the path, is moving around the binormal direction (Y-axis) with angular velocity $\Omega_{\rm D}(t) = \lambda V_0/V$.

\vspace{3mm}
\textbf{IV.} $s_{\rm \tau} \neq 0$, $s_{\rm n} = 0$, $s_{\rm b} \neq 0$. Just as in the case \textbf{III} from eq.~(\ref{2.18}) we have $V\ddot{\Theta} + \dot{V}\dot{\Theta} = 0$, $\dot{\Theta} = \lambda V_0/V$. Then eq.~(\ref{2.21}) reduces to the equation for function $\Theta(t)$
\begin{equation}
 \frac{d}{dt} \left [ \frac{\dddot{\Theta}}{\dot{\Theta}^3} - \frac{2 \ddot{\Theta}^2}{\dot{\Theta}^4} - \frac{\Omega^2_0}{\dot{\Theta}^2} + \ln{\dot{\Theta}} \right ] = \frac{2 E_Z}{\varsigma s_{\rm b} \lambda \sqrt{\dot{\Theta}}} \frac{d}{dt} \frac{\sin{\Theta}}{\sqrt{\dot{\Theta}}} \; , \label{4.20}
\end{equation}
which has solution if we put
\begin{equation}
 \dot{\Theta} = i \Omega_0 \sin{\Theta} \; , \quad \lambda = - \frac{i E_Z}{\varsigma s_{\rm b} \Omega_0} \; . \label{4.21}
\end{equation}
Then
\begin{equation}
 \cos{\Theta(t)} = \frac{\sin{\Omega_0 t} + i \cos{\Theta_0} \cos{\Omega_0 t}}{\cos{\Theta_0} \sin{\Omega_0 t} + i \cos{\Omega_0 t}} = \frac{\cos{\Theta_0} - i \sin^2{\Theta_0} \sin{\Omega_0 t} \cos{\Omega_0 t}}{\cos^2{\Theta_0} \sin^2{\Omega_0 t} + \cos^2{\Omega_0 t}} \; , \label{4.22}
\end{equation}
whence it follows that eq.~(\ref{4.20}) has a real solution only if the imaginary part of eq.~(\ref{4.22}) vanishes, i.~e. $\sin^2{\Theta_0}=0$, corresponding to either $\Theta(t) = \Theta_0 = 0$ or $\Theta(t) = \Theta_0 = \pi$. As a result eq.~(\ref{2.17}) takes the form $\ddot{V} + \Omega^2_0 V = 0$ with solution
\begin{equation}
 V(t) = V_0 \cos{(\Omega_0 t + \varphi_0)} \; . \label{4.23}
\end{equation}
Therefore,
\begin{equation}
 \mathbf{V}(t) = V_0 \cos{\Theta_0} \cos{(\Omega_0 t + \varphi_0)} \mathbf{e}_Z \; , \label{4.24}
\end{equation}
\begin{equation}
 \mathbf{R}(t) = \mathbf{R}(0) + \frac{V_0 \cos{\Theta_0}}{\Omega_0} \left [ \sin{(\Omega_0 t + \varphi_0)} - \sin{\varphi_0} \right ] \mathbf{e}_Z \; . \label{4.25}
\end{equation}

Eq.~(2.16) gives
\begin{equation}
 \frac{du}{dV} = m_0 V - E_Z t \cos{\Theta_0} = m_0 V - \frac{E_Z \cos{\Theta_0}}{\Omega_0} \left [ \arccos{(V/V_0)} - \varphi_0 \right ] \; , \label{4.26}
\end{equation}
\begin{equation}
 u(V) = \frac{m_0 V^2}{2} - \frac{E_Z \cos{\Theta_0}}{\Omega_0} V \left [ \arccos{(V/V_0)} - \varphi_0 \right ] + \frac{E_Z \cos{\Theta_0}}{\Omega_0} \sqrt{V^2_0 - V^2} \; , \label{4.27}
\end{equation}
\begin{equation}
 \gamma(V) = 1 - \frac{du}{m_0 V dV} = \frac{E_Z \cos{\Theta_0}}{m_0 \Omega_0 V} \left [ \arccos{(V/V_0)} - \varphi_0 \right ] \; . \label{4.28}
\end{equation}

So, the particle with spin $\mathbf{s} = s_{\rm b} \mathbf{e}_Y + s_{\rm \tau} \cos{\Theta_0} \mathbf{e}_Z$, flying into electric field $\mathbf{E}$, is moving parallel ($\Theta_0=0$) or antiparallel ($\Theta_0=\pi$) to the field, making thus oscillate along the direction of the field with an amplitude $r_0 = V_0 / \Omega_0$ about the initial position of the particle.

\vspace{3mm}
\textbf{V.} $s_{\rm \tau} = 0$, $s_{\rm n} \neq 0$, $s_{\rm b} = 0$. Just as in the previous case eq.~(\ref{2.18}) has solution (\ref{4.23}), eq.~(\ref{2.21}) reduces to eq.~(\ref{3.15}) with solution (\ref{3.16}), and eq.~(\ref{2.17}) leads to eqs.~(\ref{3.18}), (\ref{3.19}). By expressing the time $t$ from eq.~(\ref{4.23}) and substituting it into eq.~(\ref{3.18}), we obtain the equation for function $u(V)$
\begin{equation}
 \frac{du}{dV} = m_0 V + \frac{E_Z}{\lambda} \sqrt{1 + \left ( \cot{\Theta_0} + \frac{\lambda}{\Omega_0} \varphi_0 - \frac{\lambda}{\Omega_0} \arccos{\frac{V}{V_0}} \right )^2} \; , \label{4.29}
\end{equation}
from which we find
\begin{equation}
 u(V) = \frac{m_0 V^2}{2} + \frac{E_Z}{\lambda} \int^V_0 {\sqrt{1 + \left ( \cot{\Theta_0} + \frac{\lambda}{\Omega_0} \varphi_0 - \frac{\lambda}{\Omega_0} \arccos{\frac{V}{V_0}} \right )^2} dV} \; , \label{4.30}
\end{equation}
\begin{equation}
 \gamma(V) = 1 - \frac{du}{m_0 V dV} = - \frac{E_Z}{m_0 \Omega_0 V} {\sqrt{\frac{\Omega^2_0}{\lambda^2} + \left ( \arccos{\frac{V}{V_0}} - \varphi_0 - \frac{\Omega_0}{\lambda} \cot{\Theta_0} \right )^2}} \; . \label{4.31}
\end{equation}

The velocity and radius vector of the particle are given by
\begin{equation}
 \mathbf{V}(t) = \frac{V_0 \cos{(\Omega_0 t + \varphi_0)}}{\sqrt{1 + (\cot{\Theta_0} - \lambda t)^2}} \left [ \cos{\Phi_0} \mathbf{e}_X + \sin{\Phi_0} \mathbf{e}_Y + (\cot{\Theta_0} - \lambda t) \mathbf{e}_Z \right ] \; , \label{4.32}
\end{equation}
\begin{equation}
 \mathbf{R}(t) = \mathbf{R}(0) + \rho_0 \sin{\Theta_0} I_1(t) \left [ \cos{\Phi_0} \mathbf{e}_X + \sin{\Phi_0} \mathbf{e}_Y \right ] + \rho_0 \left [ \cos{\Theta_0} I_1(t) + I_2(t) \right ] \mathbf{e}_Z \; , \label{4.33}
\end{equation}
where
\begin{equation}
 I_1(t) = \int^t_0 \frac{\cos{(\Omega_0 t + \varphi_0)} dt}{\sqrt{t^2 + 2 \tau_0 t \cos{\Theta_0} + \tau^2_0}} \; , \quad I_2(t) = \frac{1}{\tau_0} \int^t_0 \frac{t \cos{(\Omega_0 t + \varphi_0)} dt}{\sqrt{t^2 + 2 \tau_0 t \cos{\Theta_0} + \tau^2_0}} \; , \label{4.34}
\end{equation}
\begin{equation}
 \tau_0 = \tau(V_0) = - \frac{1}{\lambda \sin{\Theta_0}} = \frac{m_0 \gamma_0 V_0}{E_Z} \; , \quad \rho_0 = V_0 \tau_0 = \frac{m_0 \gamma_0 V^2_0}{E_Z} \; . \label{4.35}
\end{equation}

Unfortunately, eqs.~(\ref{4.30}) and (\ref{4.33}) contain the integrals (\ref{4.34}), which cannot be expressed in terms of known functions, except for the case $\Omega_0=0$, when
\begin{equation}
 I_1(t) = \cos{\varphi_0} \ln{\frac{t + \tau_0 \cos{\Theta_0} + \sqrt{t^2 + 2 \tau_0 t \cos{\Theta_0} + \tau^2_0 (1 + \cos^2{\Theta_0})}}{\tau_0 \left [ \cos{\Theta_0} + \sqrt{1 + \cos^2{\Theta_0}} \right ]}} \; , \label{4.36}
\end{equation}
\begin{equation}
 I_2(t) = \frac{\cos{\varphi_0}}{\tau_0} \left [ \sqrt{t^2 + 2 \tau_0 t \cos{\Theta_0} + \tau^2_0} - \tau_0 \right ] - I_1(t) \cos{\Theta_0} \; . \label{4.37}
\end{equation}

Spin
\begin{equation}
 \mathbf{s} = s_{\rm n} \mathbf{e}_{\rm n} = \frac{s_{\rm n}}{\sqrt{1 + (\cot{\Theta_0} - \lambda t)}} \left [ (\cot{\Theta_0} - \lambda t) \mathbf{e}_X - \mathbf{e}_Z \right ] \; \label{4.38}
\end{equation}
of the particle flying into electric field $\mathbf{E}$ under the angle $\Theta_0$ precesses with angular velocity
\begin{equation}
 \mathbf{\Omega}_{\rm D} = \dot{\Theta} \mathbf{e}_{\rm b} = - \frac{E_Z \sin{\Theta_0}}{m_0 V_0 \gamma_0 \left [ \sin^2{\Theta_0} + (\cos{\Theta_0} - \lambda t \sin{\Theta_0})^2 \right ]} \mathbf{e}_{\rm b}  \label{4.39}
\end{equation}
around the binormal direction $\mathbf{e}_{\rm b}$.

\vspace{3mm}
\textbf{VI.} $s_{\rm \tau} = 0$, $s_{\rm n} \neq 0$, $s_{\rm b} \neq 0$. Just as in the previous case eq.~(\ref{2.18}) has solution (\ref{4.23}), eq.~(\ref{2.21}) reduces to the equation for $\Theta(t)$
\begin{equation}
 \frac{2}{ \dot{\Theta} \sqrt{\dot{\Theta}}} \frac{d}{dt} \frac{\sin{\Theta}}{\sqrt{\dot{\Theta}}} = \frac{\varsigma s_{\rm b} \Omega_0 V_0}{E_Z} \sin{(\Omega_0 t + \Theta_0)} \; , \label{4.40}
\end{equation}
the solution of which is a linear function of time
\begin{equation}
 \dot{\Theta} = \Omega_0 \; , \quad \Theta(t) = \Omega_0 t + \Theta_0 + \pi/2 \; , \label{4.41}
\end{equation}
where
\begin{equation}
 \Omega^2_0 = - \frac{2E_Z}{\varsigma s_{\rm b} V_0} > 0 \; . \label{4.42}
\end{equation}

Substituting eq.~(\ref{4.41}) into eq.~(\ref{2.17}) gives the equation for function gives the equation for function $u(V)$
\begin{equation}
 (m_0 - \varsigma s_{\rm b} \Omega_0) V - \frac{du}{dV} = - \frac{E_Z}{\Omega_0} \cos{(\Omega_0 t + \Theta_0)} = - \frac{E_Z}{\Omega_0 V_0} V \; , \label{4.43}
\end{equation}
from which we find
\begin{equation}
 u(V) = \frac{1}{2} \left ( m_0 - \varsigma s_{\rm b} \Omega_0 + \frac{E_Z}{\Omega_0 V_0} \right ) V^2 \; , \label{4.44}
\end{equation}
\begin{equation}
 \gamma(V) = 1 - \frac{du}{m_0 V dV} = \frac{\varsigma s_{\rm b} \Omega_0}{m_0} - \frac{E_Z}{m_0 \Omega_0 V_0} = \mathrm{const} \; . \label{4.45}
\end{equation}

Note that when $s_{\rm b} = 0$ the expression (\ref{4.44}) is a limiting case of eq.~(\ref{4.30}), when $V<<V_0$ and $\Theta_0 = \pi/2$.

Thus,
\begin{equation}
 \mathbf{V}(t) = V_0 \cos{(\Omega_0 t + \Theta_0)} \left [ \cos{(\Omega_0 t + \Theta_0)} (\cos{\Phi_0} \mathbf{e}_X + \sin{\Phi_0} \mathbf{e}_Y) - \sin{(\Omega_0 t + \Theta_0)} \mathbf{e}_Z \right ] \; , \label{4.46}
\end{equation}
from which we have equation of trajectory (Fig. 1)
\begin{equation}
\begin{split}
 \mathbf{R}(t) &= \mathbf{R}(0) + \frac{V_0}{4 \Omega_0} \left [ 2 \Omega_0 t + \sin{2(\Omega_0 t + \Theta_0)} - \sin{2\Theta_0} \right ] (\cos{\Phi_0} \mathbf{e}_X + \sin{\Phi_0} \mathbf{e}_Y) + \\ {} & \quad + \frac{V_0}{4 \Omega_0} \left [ \cos{2(\Omega_0 t + \Theta_0)} - \cos{2\Theta_0} \right ] \mathbf{e}_Z \; . \label{4.47}
\end{split}
\end{equation}

\begin{center}
\includegraphics[scale=0.9,keepaspectratio,draft=false]{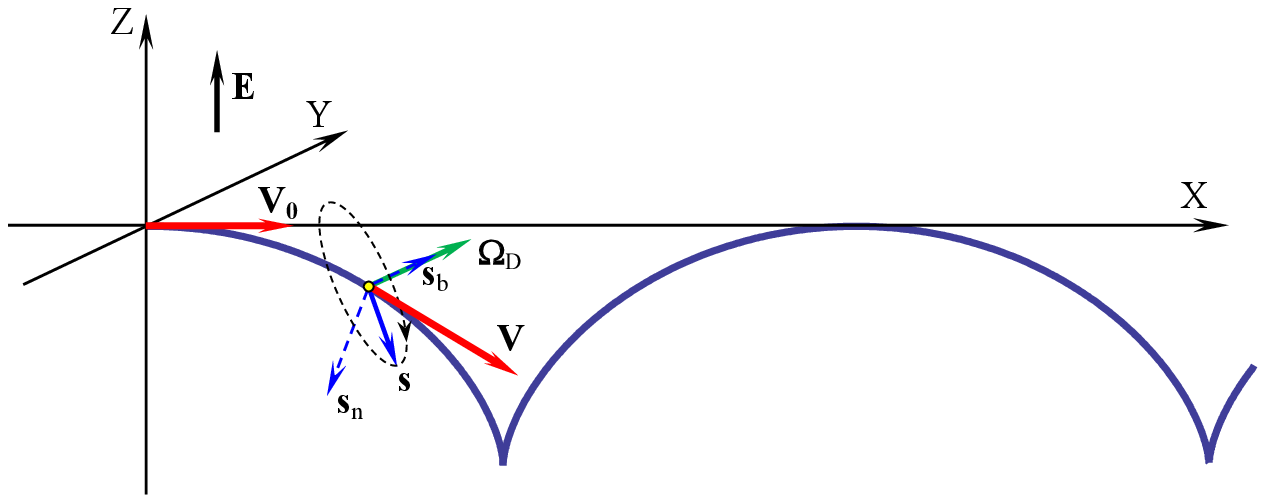}
\end{center}
\vspace{-5mm}
\centerline{Fig. 1. An example of trajectory (\ref{4.47}) of the electron in electric field at $\Phi_0 = 0$, $\Theta_0 = 0$}
\vspace{5mm}

According to eq.~(\ref{2.23}) spin of the particle in the coordinate system, where $\Phi_0=0$,
\begin{equation}
 \mathbf{s} = - s_{\rm n} \sin{(\Omega_0 t + \Theta_0)} \mathbf{e}_X + s_{\rm b} \mathbf{e}_Y - s_{\rm n} \cos{(\Omega_0 t + \Theta_0)} \mathbf{e}_Z \; \label{4.48}
\end{equation}
is perpendicular to the direction of motion and precesses around the binormal direction (Y-axis) with angular velocity $\Omega_{\rm D} = \Omega_0$. As seen in Fig. 1, the electron trajectory (or any charged particle) has a rather strange ``jumping'' form.

\vspace{3mm}
\textbf{VII.} $s_{\rm \tau} \neq 0$, $s_{\rm n} \neq 0$, $s_{\rm b} = 0$. Equations (\ref{2.16})-(\ref{2.17}) are reduced to eq.~(\ref{3.15}) with solution (\ref{3.16}). Substituting it into eq.~(\ref{2.18}), we get the equation for function $V(t)$
\begin{equation}
  \frac{d^2 V}{d x^2} + \frac{d(f V)}{dx} + \frac{\Omega^2_0}{\lambda^2} V = 0 \; , \label{4.49}
\end{equation}
with
\begin{equation}
  x = \cot{\Theta(t)} = \cot{\Theta_0} - \lambda t \; , \quad f(x) = \frac{s_{\rm \tau}}{s_{\rm n} (1+x^2)} \; . \label{4.50}
\end{equation}

Trajectory equation according to eqs.~(\ref{2.13}) and (\ref{3.17}) has the form
\begin{equation}
 \mathbf{R}(t) = \mathbf{R}(0) + R_1(t) (\cos{\Phi_0} \mathbf{e}_X + \sin{\Phi_0} \mathbf{e}_Y) + R_2(t) \mathbf{e}_Z \; , \label{4.51}
\end{equation}
where
\begin{equation}
 R_1(t) = \int^t_0 \frac{V(t)dt}{\sqrt{1 + (\cot{\Theta_0} - \lambda t)^2}} \; , \quad R_2(t) = \int^t_0 \frac{V(t) (\cot{\Theta_0} - \lambda t) dt}{\sqrt{1 + (\cot{\Theta_0} - \lambda t)^2}} \; . \label{4.52}
\end{equation}

In general, integrals (\ref{4.52}) cannot be expressed in terms of known functions. However, if we assume that spin $\mathbf{s} = s_{\rm \tau} \mathbf{e}_{\rm \tau} + s_{\rm n} \mathbf{e}_{\rm n}$ precesses with constant angular velocity $\Omega_{\rm D}$, it is necessary to put $\Theta(t)=\Omega_{\rm D} t + \Theta_0$. Then eq.~(\ref{2.18}) is an equation of damped oscillations, which has solution
\begin{equation}
 V(t) = \left \{
          \begin{array}{ll}
            V_0 e^{k_1 \Omega_{\rm D} t} \cos{(\Omega t + \varphi_0)} \; , & \hbox{if} \quad \Omega^2 = \Omega^2_0 (1 - k^2_1) > 0 \; ; \\
            V_1 e^{\tilde{\Omega}_{+} t} + V_2 e^{\tilde{\Omega}_{-} t} \; , & \hbox{if} \quad \Omega^2 = - \tilde{\Omega}^2 = \Omega^2_0 (1 - k^2_1) \leq 0 \; .
          \end{array}
        \right.
 \label{4.53}
\end{equation}

Trajectory equation looks like
\begin{equation}
\begin{split}
 \mathbf{R}(t) = \mathbf{R}(0) &+ \frac{V_0 e^{k_1 \Omega_{\rm 0} t} \cos{\Phi_0}}{2 \Omega_0} \left [ \frac{k_1 \sin{\Psi_{-}} - (k_2 - \sqrt{1 - k^2_1}) \cos{\Psi_{-}}}{(k_2 - \sqrt{1 - k^2_1})^2 + k^2_1} + \right. \\ {} &+ \left. \frac{k_1 \sin{\Psi_{+}} - (k_2 + \sqrt{1 - k^2_1}) \cos{\Psi_{+}}}{(k_2 + \sqrt{1 - k^2_1})^2 + k^2_1} \right ] \mathbf{e}_X + \\ {} &+ \frac{V_0 e^{k_1 \Omega_{\rm 0} t} \sin{\Phi_0}}{2 \Omega_0} \left [ \frac{k_1 \sin{\Psi_{-}} - (k_2 - \sqrt{1 - k^2_1}) \cos{\Psi_{-}}}{(k_2 - \sqrt{1 - k^2_1})^2 + k^2_1} + \right. \\ {} &+ \left. \frac{k_1 \sin{\Psi_{+}} - (k_2 + \sqrt{1 - k^2_1}) \cos{\Psi_{+}}}{(k_2 + \sqrt{1 - k^2_1})^2 + k^2_1} \right ] \mathbf{e}_Y + \\ {} &+ \frac{V_0 e^{k_1 \Omega_{\rm 0} t}}{2 \Omega_0} \left [ \frac{k_1 \cos{\Psi_{-}} + (k_2 - \sqrt{1 - k^2_1}) \sin{\Psi_{-}}}{(k_2 - \sqrt{1 - k^2_1})^2 + k^2_1} + \right. \\ {} &+ \left. \frac{k_1 \cos{\Psi_{+}} + (k_2 + \sqrt{1 - k^2_1}) \sin{\Psi_{+}}}{(k_2 + \sqrt{1 - k^2_1})^2 + k^2_1} \right ] \mathbf{e}_Z \; , \label{4.54}
\end{split}
\end{equation}
where
\begin{equation}
 k_1 = \cfrac{s_{\rm \tau}}{2s_{\rm n}} \cfrac{\Omega_{\rm D}}{\Omega_0} = k_2 \cfrac{s_{\rm \tau}}{2s_{\rm n}} \; , \quad \Omega_{\rm D} = k_2 \Omega_0 \; , \label{4.55}
\end{equation}
\begin{equation}
 \Psi_{\pm}(t) = \Omega_{\pm} t + \Phi_{\rm D} \pm \varphi_0 \; , \quad \Omega_{\pm} = \Omega_{\rm D} \pm \Omega = (k_2 - \sqrt{1 - k^2_1}) \Omega_0 \; , \label{4.56}
\end{equation}
if $\Omega^2 = \Omega^2_0 (1 - k^2_2) > 0$, or

\begin{equation}
\begin{split}
 \mathbf{R}(t) &= \mathbf{R}(0) + \\ {} &+ \frac{V_1 e^{\tilde{\Omega}_{+} t}}{\tilde{\Omega}^2_{+} + \Omega^2_{\rm D}} \left[ \tilde{\Omega}_{+} \sin{(\Omega_{\rm D} t + \Theta_0)} - \Omega_{\rm D} \cos{(\Omega_{\rm D} t + \Theta_0)} \right ] (\cos{\Phi_0} \mathbf{e}_X + \sin{\Phi_0} \mathbf{e}_Y) + \\ {} &+ \frac{V_2 e^{\tilde{\Omega}_{-} t}}{\tilde{\Omega}^2_{-} + \Omega^2_{\rm D}} \left[ \tilde{\Omega}_{-} \sin{(\Omega_{\rm D} t + \Theta_0)} - \Omega_{\rm D} \cos{(\Omega_{\rm D} t + \Theta_0)} \right ] (\cos{\Phi_0} \mathbf{e}_X + \sin{\Phi_0} \mathbf{e}_Y) + \\ {} &+ \frac{V_1 e^{\tilde{\Omega}_{+} t}}{\tilde{\Omega}^2_{+} + \Omega^2_{\rm D}} \left[ \tilde{\Omega}_{+} \cos{(\Omega_{\rm D} t + \Theta_0)} + \Omega_{\rm D} \sin{(\Omega_{\rm D} t + \Theta_0)} \right ] \mathbf{e}_Z + \\ {} &+ \frac{V_2 e^{\tilde{\Omega}_{-} t}}{\tilde{\Omega}^2_{-} + \Omega^2_{\rm D}} \left[ \tilde{\Omega}_{-} \cos{(\Omega_{\rm D} t + \Theta_0)} + \Omega_{\rm D} \sin{(\Omega_{\rm D} t + \Theta_0)} \right ] \mathbf{e}_Z \; , \label{4.57}
\end{split}
\end{equation}
where
\begin{equation}
\tilde{\Omega}_{\pm} = (k_1 - \sqrt{k^2_1 -1}) \Omega_0 \; , \label{4.58}
\end{equation}
if $\Omega^2 = - \tilde{\Omega}^2 = \Omega^2_0 (1 - k^2_1) \leq 0$.

Trajectories (\ref{4.54}) are rather complicated stellate type curves, whereas (\ref{4.57}) are smooth curves and similar to observable trajectories.

\vspace{3mm}
\textbf{VIII.} $s_{\rm \tau} \neq 0$, $s_{\rm n} \neq 0$, $s_{\rm b} \neq 0$. Note that all cases \textbf{I-III}, \textbf{V} and \textbf{VII} satisfy to eq.~(\ref{3.15}), for the case \textbf{IV} we have eq.~ (\ref{4.21}), and eq.~(\ref{4.41}) for the case \textbf{VI}. All these equations can be combined in single equation, corresponding to the general case under consideration, as
\begin{equation}
 \dot{\Theta} = \Omega_0 (\alpha_1 \sin^2{\Theta} + \alpha_2 \sin{\Theta} + 1) \; , \label{4.59}
\end{equation}
whence it follows
\begin{equation}
 \ddot{\Theta} = \Omega_0 \dot{\Theta} (2\alpha_1 \sin{\Theta} + \alpha_2) \cos{\Theta} \; , \label{4.60}
\end{equation}
\begin{equation}
 \int^{\Theta}_{\Theta_0} \frac{d\Theta}{\alpha_1 \sin^2{\Theta}+ \alpha_2 \sin{\Theta} +1} = \Omega_0 t \; . \label{4.61}
\end{equation}

Integration of eq.~(\ref{4.61}) gives
\begin{equation}
\begin{split}
 \Omega_0 t \sqrt{\alpha^2_2 - 4 \alpha_1} &= \frac{1}{1 - x^2_1} \ln{\frac{x_1 x + 1 + \sqrt{(1 - x^2_1)(1 - x^2)}}{x + x_1}} - \\ {} & \:-\frac{1}{1 - x^2_2} \ln{\frac{x_2 x + 1 + \sqrt{(1 - x^2_2)(1 - x^2)}}{x + x_2}} \; , \label{4.62}
\end{split}
\end{equation}
where $x=\sin{\Theta}>0$, $\alpha^2_2 > 4 \alpha_1$,
\begin{equation}
 x_{1,2} = \frac{\alpha_2 \pm \sqrt{\alpha^2_2 - 4 \alpha_1}}{2 \alpha_1} \; . \label{4.63}
\end{equation}

For the cases $\alpha^2_2 < 4 \alpha_1$ and $\alpha^2_2 = 4 \alpha_1$ the integral (\ref{4.61}) is also expressed in terms of elementary functions, but the final expressions are too bulky to introduce them here.

As concerns the expression (\ref{4.62}), then, in spite of its relative simplicity, its transcendence does not allows to find the explicit dependence of the function $\Theta(t)$ on time, which would make it possible to find $V(t)$ from eq.~(\ref{2.18}) and $u(V)$ from eqs.~(\ref{2.19})-(\ref{2.20}).

In the simplest case $\alpha_1=\alpha_2=0$ we find $\Theta(t)=\Omega_0 t + \Theta_0$; eq.~(\ref{2.18}) becomes the equation of damped oscillations. As a result, the trajectory equation looks like eq.~(\ref{4.54}), if $4s^2_{\rm n}>s^2_{\rm \tau}$, or (\ref{4.57}), if $4s^2_{\rm n} \leq s^2_{\rm \tau}$, provided $\Omega_{\rm D}=\Omega_0$.

\section{Acceleration of electrons by an electric field}

\renewcommand{\thesection}{\arabic{section}}
\renewcommand{\theequation}{\thesection.\arabic{equation}}
\setcounter{section}{5}
\setcounter{equation}{0}

Let a charged particle flies into electric field $\mathbf{E}$, directed along the Z-axis (Sec. 2) with initial velocity $\mathbf{V}_0 = e V_0 \mathbf{e}_Z$, where $V_0>0$, $e = \cos{\Theta_0} = \pm 1$, which corresponds to the motion either along ($\Theta_0=0$) or against ($\Theta_0=\pi$) the field. Then in classical and relativistic case we find from eqs.~(\ref{3.6}) and (\ref{3.13})
\begin{equation}
 Z(t) = Z(0) + e V_0 t + E_Z t^2 / 2m_0 \; , \label{5.1}
\end{equation}
\begin{equation}
 Z(t) = Z(0) + c \tau_0 \gamma_0 \left [ \sqrt{\gamma^{-2}_0 \frac{t^2}{\tau^2_0} + 2 e \beta_0 \gamma^{-1}_0 \frac{t}{\tau_0}+1} -1 \right ] \; , \label{5.2}
\end{equation}
respectively, where
\begin{equation}
 \tau_0 = \frac{1}{\lambda_0} = \frac{m_0 c}{E_Z} \; , \quad \beta_0 = \frac{V_0}{c} \; , \quad \gamma_0 = \gamma_{\rm L}(V_0) = (1 - \beta^2_0)^{-1/2} \; . \label{5.3}
\end{equation}

For the case \textbf{III} ($s_{\rm \tau} \neq 0$, $s_{\rm n} = s_{\rm b} = 0$) it follows from eqs.~(\ref{2.19})-(\ref{2.20})
\begin{equation}
 m_0 V \gamma(V) \sin{\Theta} = C_1 \; , \quad m_0 V \gamma(V) \cos{\Theta} = E_Z t + C_2 \; , \label{5.4}
\end{equation}
where $\gamma(V)$ is given by the first equality in eq.~(\ref{4.16}).

Since, according to eq.~(\ref{2.9})
\begin{equation}
  \mathbf{V}(t) = V(t) \left[\sin{\Theta} \mathbf{e}_X + \cos{\Theta} \mathbf{e}_Z \right] = \frac{1}{m_0 \gamma(V)} \left[C_1 \mathbf{e}_X + (E_Z t + C_2) \mathbf{e}_Z \right] \; , \label{5.5}
\end{equation}
\begin{equation}
  \mathbf{V}_0 = V_0 \left[\sin{\Theta}_0 \mathbf{e}_X + \cos{\Theta}_0 \mathbf{e}_Z \right] = \frac{1}{m_0 \gamma(V_0)} \left[C_1 \mathbf{e}_X + C_2 \mathbf{e}_Z \right] = V_0 \cos{\Theta}_0 \mathbf{e}_Z \; , \label{5.6}
\end{equation}
then we have $C_1=0$, $C_2 = e m_0 V_0 \gamma(V_0)$. Therefore eqs.~(\ref{5.4}) and (\ref{5.5}) give $\Theta = \Theta_0 = 0,\pi$ and
\begin{equation}
  \gamma(V) = 1 - \frac{du}{m_0 V dV} = \gamma(V_0) + \frac{e E_Z t}{m_0 V} = \gamma(V_0) + \frac{e t}{\beta \tau_0}\; , \label{5.7}
\end{equation}
where $\beta = V/c$. This relation does not allow to find the explicit dependence of $\gamma(V)$, because from eq.~(\ref{5.4}) function $V(t)$ cannot be determined. While in the relativistic case the function $\gamma(V)$ is the Lorentz factor, eq.~(\ref{3.3}), then here for its definition one needs to know the form of function $u(V)$. Then the velocity is found from eq.~(\ref{5.5}), or
\begin{equation}
  \mathbf{V}(t) = \frac{e m_0 V_0 \gamma(V_0) + E_Z t}{m_0 \gamma(V)} \mathbf{e}_Z = c \frac{e \beta_0 \gamma(V_0) + t/ \tau_0}{\gamma(V)} \mathbf{e}_Z  \; , \label{5.8}
\end{equation}
from where $Z(t)$ may be found by integration.

In general, if $V_0$, and $V_1$ are initial and final velocity of the particle, then we have from the law (\ref{2.6}) of energy conservation
\begin{equation}
  \frac{m_0 (V^2_1 - V^2_0)}{2} + V^2_0 \left. \frac{d(u/V)}{dV} \right |_{V=V_0} - V^2_1 \left. \frac{d(u/V)}{dV} \right |_{V=V_1} + \Delta \varphi = 0 \; , \label{5.9}
\end{equation}
where $\Delta \varphi = \varphi_1 - \varphi_0$ is accelerating potential difference. For the classical particle we get the obvious relation at $u=0$, and for relativistic one we find from eqs.~(\ref{3.4}) and (\ref{5.9})
\begin{equation}
  \frac{m_0 c^2}{\sqrt{1 - V^2_0 / c^2}} - \frac{m_0 c^2}{\sqrt{1 - V^2_1 / c^2}} + \Delta \varphi = 0 \; . \label{5.10}
\end{equation}

Thus, the relation (\ref{5.9}) shows that the acceleration of particle by the electric field is determined only by the form of function  $u(v)$, but not its spin.

\section{Deflection of electrons in a constant electric field}

\renewcommand{\thesection}{\arabic{section}}
\renewcommand{\theequation}{\thesection.\arabic{equation}}
\setcounter{section}{6}
\setcounter{equation}{0}

We assume, as in the papers of Kaufmann (\cite{Kau6}, S. 525) and Planck~\cite{Pla1}, that the electron is emitted by the radiation source, whose coordinates are $\textbf{R}_0 = (X_0,Y_0,Z_0) = (0,0,0)$ in the direction of the X-axis, then passes through the diaphragm aperture, whose coordinates are $\textbf{R}_1 = (X_1,Y_1,Z_1) = (X_1,0,0)$, to the photographic plate, on which its coordinates become $\textbf{R}_2 = (X_2,Y_2,Z_2)$. The velocity of the electron, emitted from the diaphragm, we take as the initial one, and initial time is denoted by $t_0$. According to \cite{Tar2} free electron has a longitudinal polarization of negative helicity and is moving with the velocity $\mathbf{V} = \mathbf{V}_{(\rm K')} + \mathbf{v}$, where $\mathbf{V}_{(\rm K')} = V_0 \mathbf{e}_X = (V_0,0,0) = \mathrm{const}$, and $\mathbf{v}$ is given by eq.~(\ref{3.10}), or eq.~(\ref{3.19}) or $\mathbf{v} = \mathbf{0}$ (with cyclic substitute $Z \rightarrow X$, $X \rightarrow Y$, $Y \rightarrow Z$), that corresponds to the Lorentz electron. Consequently, if $\mathbf{v} \neq \mathbf{0}$, then when approaching the aperture only those electrons will pass from it, that have vanished transverse component of the velocity, $\mathbf{v}(t_0) = \mathbf{0}$, i.~e. at
\begin{equation}
 \Omega_0 t_0 + \varphi_0 = (2k+1) \pi / 2 \; \label{6.1}
\end{equation}
for eq.~(\ref{3.10}) from \cite{Tar2} or
\begin{equation}
 \frac{F \Omega_{\rm D}}{\Omega^2_{\rm D} + \Omega^2_0} + v_0 \cos{(\chi t_0 + \varphi_0)} = 0 \; \label{6.2}
\end{equation}
for eq.~(\ref{3.19}) from \cite{Tar2}. In this case we have $X_1 = V_0 t_0$. After that the character of the electron motion will be determined by a field in which it moves.

Let in the chosen coordinate system a constant electric force is directed along the Z-axis, $\mathbf{E} = E_Z \mathbf{e}_Z$, $E_Z > 0$. Accordingly, the angle between the initial velocity and the field $\mathbf{E}$ is $\Theta = \Theta_0 = \pi/2$. In addition, we assume $\Phi = \Phi_0 =0$ that corresponds to the electron motion in the XZ-plane. Then electric deflection $Z_2$ of the electron is determined from eq.~(\ref{3.6}) for classical electron, or from eq.~(\ref{3.13}) for the Lorentz-Einstein electron, or from eq.~(\ref{4.19}) for spinning electron. In these equations $t=t_0$, so that $\mathbf{R}(0) = \mathbf{R}(t_0) = \mathbf{R}_1 = (V_0 t_0,0,0)$, as well as $Y_2(t)=0$. Without loss of generality the origin of coordinate system can be placed on the diaphragm, if putting $t_0=0$. The distance that the electron passes along the X-axis, denoted by $L = X_2 - X_1$.

We use the dimensionless coordinates of the electron at a time $t$, $x(t)=X/L$, $z(t)=Z/L$, as well as dimensionless constants (\ref{5.3}). Then from eq.~(\ref{3.6}) for classical electron we obtain
\begin{equation}
 x(t) = \frac{V_0 t}{L} = \frac{c \tau_0}{L} \beta_0 \frac{t}{\tau_0} \; , \quad z(t) = \frac{E_Z t^2}{m_0 L}= \frac{c \tau_0}{L} \frac{t^2}{\tau^2_0} \; , \label{6.3}
\end{equation}
The corresponding curves in Figs. 2-3 are shown in red color.

From eq.~(\ref{3.13}), where $t_1=0$, $\gamma_0 = \gamma_{\rm L}(0)$, for the Lorentz-Einstein electron it follows (blue curves in Figs. 2-3)
\begin{equation}
 \left \{
\begin{array}{c}
\begin{aligned}
           x(t)& = \cfrac{c \tau_0}{L} \beta_0 \gamma_0 \ln \left [1 + \cfrac{\gamma^{-1}_0 t/\tau_0}{\sqrt{\gamma^{-2}_0 (t/\tau_0)^2 + 1}} \right ] \; , \\
           z(t)& = \cfrac{c \tau_0}{L} \gamma_0 \left [ \sqrt{\gamma^{-2}_0 (t/\tau_0)^2 + 1} - 1 \right ] \; .
\end{aligned}
         \end{array}
\right.
  \label{6.4}
\end{equation}

Finally, assuming $\lambda=\tau^{-1}_0$ in eq.~(\ref{4.19}), we find (green curves in Figs. 2-3)
\begin{equation}
 \left \{
\begin{array}{c}
\begin{aligned}
           x(t)& = \cfrac{c \tau_0}{L} \cfrac{\beta_0}{2} \left [\cfrac{t}{\tau_0} \sqrt{1 + \cfrac{t^2}{\tau^2_0}} - \ln \left ( \sqrt{1 + \cfrac{t^2}{\tau^2_0}} - \cfrac{t}{\tau_0} \right ) \right ] \; , \\
           z(t)& = \cfrac{c \tau_0}{L} \cfrac{\beta_0}{3} \left [ 1 - \left ( 1 + \cfrac{t^2}{\tau^2_0} \right )^{3/2}\right ] \; .
\end{aligned}
         \end{array}
\right.
  \label{6.5}
\end{equation}

On the curves shown in Fig.~2 and representing electrostatic deflection in these three cases, one can see that the deflection of the Lorentz-Einstein relativistic electron is faster at low speeds ($\beta_0=0,001$) (Fig.~2a), and slower at higher speeds ($\beta_0=0,99$) (Fig.~2b) than deflection of spinning electron. Fig.~3 shows corresponding curves near the value $\beta_0 \approx 0,55$, when the trajectory of spinning electron becomes similar to the trajectory of relativistic electron. This means that the actual deflection of the electron can be described by eqs.~(\ref{6.5}) with another initial velocity $c \beta_{\rm s}$ and time of motion $t_{\rm s}$, whose correlation with $\beta_0$, $t_{\rm L}$ of the Lorentz-Einstein electron may be found from two transcendental equations by equating the coordinates (\ref{6.5}) and (\ref{6.4}):
\begin{equation}
 \beta_0 \gamma_0 \ln \left [1 + \cfrac{t_{\rm L}/\tau_0}{\sqrt{(t_{\rm L}/\tau_0)^2 + \gamma^2_0}} \right ] = \cfrac{\beta_{\rm s}}{2} \left [\cfrac{t_{\rm s}}{\tau_0} \sqrt{1 + \cfrac{t^2_{\rm s}}{\tau^2_0}} - \ln \left ( \sqrt{1 + \cfrac{t^2_{\rm s}}{\tau^2_0}} - \cfrac{t_{\rm s}}{\tau_0} \right ) \right ] \; , \label{6.6}
\end{equation}
\begin{equation}
 \sqrt{(t_{\rm L}/\tau_0)^2 + \gamma^2_0} - \gamma_0 = \cfrac{\beta_{\rm s}}{3} \left [ 1 - \left (1 + \cfrac{t^2_{\rm s}}{\tau^2_0} \right )^{3/2} \right ] \; . \label{6.7}
\end{equation}

The initial velocity of the $\beta$-electrons is measured from magnetic deflections by using $\beta$-spectrome\-ters (see, e.~g., \cite{Sieg}) and is usually calculated by the relativistic formulas. The first experiments on magnetic deflection of cathode rays were carried by Lenard \cite{Len}, and the behavior of $\beta$-rays in magnetic field have studied by Kaufmann \cite{Kau1}. Because the actual velocity may differ from that calculated by the Lorentz-Einstein theory in the next section we consider the magnetic deflection using the proposed equations for spinning electron.

\begin{center}
\includegraphics[scale=0.7,keepaspectratio,draft=false]{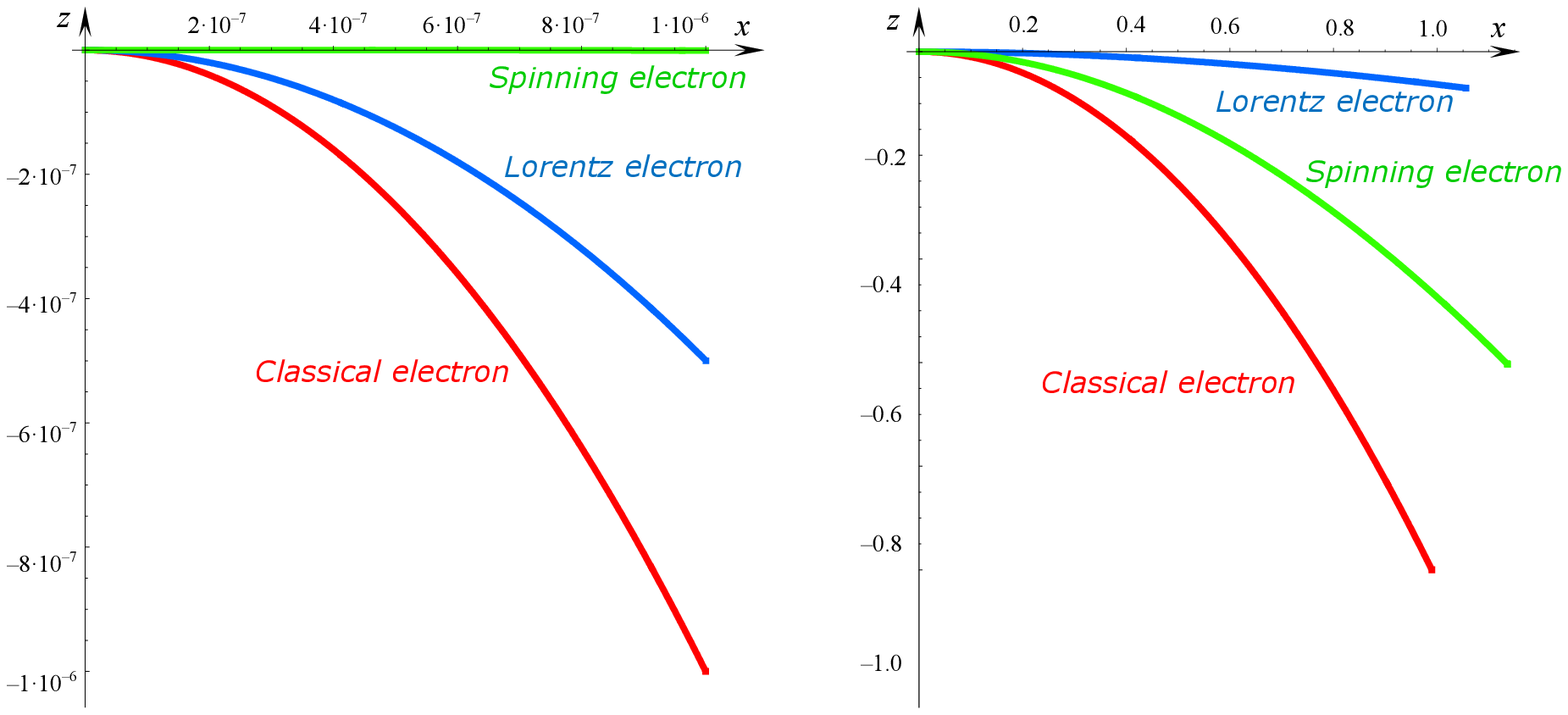}
\end{center}
\vspace{-10mm}
\centerline{a) \hspace{65mm} b)}
\vspace{3mm}
\centerline{Fig. 2. Electron trajectories at a) $\beta_0 = 0,001$; b) $\beta_0 = 0,99$}
\vspace{-5mm}
\begin{center}
\includegraphics[scale=0.7,keepaspectratio,draft=false]{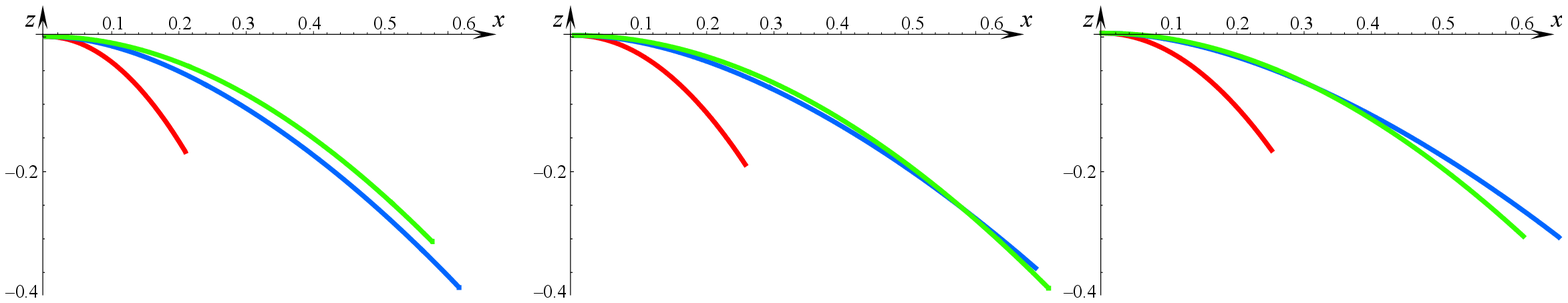}
\end{center}
\vspace{-10mm}
\centerline{a) \hspace{50mm} b) \hspace{50mm} c)}
\vspace{3mm}
\centerline{Fig. 3. Electron trajectories at a) $\beta_0 = 0,50$; b) $\beta_0 = 0,55$; c) $\beta_0 = 0,60$}
\vspace{5mm}

\section{Deflection of electrons in a constant magnetic field}

\renewcommand{\thesection}{\arabic{section}}
\renewcommand{\theequation}{\thesection.\arabic{equation}}
\setcounter{section}{7}
\setcounter{equation}{0}

Suppose, just as in the previous section, the conditions of passage of the electron through the diaphragm are satisfied. In the case of magnetic deflection the photographic plate, on which electrons fall, is located on the Y-axis, so that its coordinates are $\mathbf{R}_2 = (0,Y_2,0)$. If we choose the direction of the magnetic vector as the Z-axis, $\mathbf{B} = B_z \mathbf{e}_Z = (0,0,B_Z)$, $B_Z>0$, the motion will take place in the XY-plane. The velocity of the electron is given by eq.~(\ref{2.9}), where $\Theta = \pi/2$, i.~e.
\begin{equation}
  \mathbf{V}(t) = V(t) \mathbf{e}_{\rm \tau} = V(t) \left[\cos{\Phi(t)} \mathbf{e}_X + \sin{\Phi(t)} \mathbf{e}_Y \right] \; , \label{7.1}
\end{equation}
so that the initial velocity of the electron which flies the aperture is directed along the X-axis,
\begin{equation}
  \mathbf{V}_0 = V_0 \mathbf{e}_{\rm \tau 0} = V_0 \mathbf{e}_X = V_0 \left[\cos{\Phi_0} \mathbf{e}_X + \sin{\Phi_0} \mathbf{e}_Y \right] \; . \label{7.2}
\end{equation}

Initially, the spin (\ref{2.22}) of the electron is parallel to the velocity. This corresponds to
\begin{equation}
 s_{X0} = s_{\rm \tau}  \cos{\Phi_0} - s_{\rm n} \sin{\Phi_0} = es \; , \quad s_{Y0} = s_{\rm \tau} \sin{\Phi_0} + s_{\rm n} \cos{\Phi_0} = 0 \; , \quad s_{\rm b} = 0 \; , \label{7.3}
\end{equation}
where $\cos{\Phi_0}=e$, $\Phi_0$ is an angle between initial velocity and spin, $e=-1$ for the electron and $e=+1$ for the positron. Then eqs.~(\ref{2.22}) and (\ref{7.3}) give $s_{\rm n}=0$ and
\begin{equation}
 \mathbf{s} = es (\cos{\Phi} \mathbf{e}_X + \sin{\Phi} \mathbf{e}_Y) \; . \label{7.4}
\end{equation}

Solutions for the case $s_{\rm \tau}=s_{\rm n}=0$, obtained in \cite{Tar3}, may be regarded as asymptotic. Here we have $s_{\rm \tau} \neq 0$. Therefore, taking into account great velocity of electrons, with high probability one can be assumed that the time to the asymptotic behavior is much more than the time to reach the deflection along Y-axis. Then equation of motion (\ref{2.1}) of spinning particle in magnetic field reduce to the system
\begin{equation}
 \frac{d}{dt} \left (m_{0} V - \frac{du}{dV} \right ) = 0 \; , \label{7.5}
\end{equation}
\begin{equation}
 \left (m_{0} - \frac{du}{V dV} \right ) \dot{\Phi} + B_Z =0 \; , \label{7.6}
\end{equation}
\begin{equation}
 \frac{d}{dt}(V\dot{\Phi}) = 0 \; , \label{7.7}
\end{equation}
whence it follows
\begin{equation}
 V\dot{\Phi} = C_1 = -\frac{B_Z V_0}{m_0 \gamma(V_0)} \; , \label{7.8}
\end{equation}
\begin{equation}
 m_{0} V - \frac{du}{dV} = m_0 V \gamma(V) = C_2 \; , \label{7.9}
\end{equation}
\begin{equation}
 V = V_0 = \sqrt{-\frac{C_1 C_2}{B_Z}} = \mathrm{const} \; , \label{7.10}
\end{equation}
\begin{equation}
 \dot{\Phi} = \Omega_{\rm D} = \frac{C_1}{V_0} = -\frac{B_Z}{m_0 \gamma(V_0)} \; . \label{7.11}
\end{equation}

Substituting eq.~(\ref{7.11}) into eq.~(\ref{7.1}) we find
\begin{equation}
  \mathbf{V}(t) = V_0 \left(\cos{\Omega_{\rm D} t} \mathbf{e}_X + \sin{\Omega_{\rm D} t} \mathbf{e}_Y \right) \; . \label{7.12}
\end{equation}
Hence, the trajectory is given by
\begin{equation}
  \mathbf{R}(t) = \mathbf{R}(0) + \frac{V_0}{\Omega_{\rm D}} \left[\sin{\Omega_{\rm D} t} \mathbf{e}_X + (1 - \cos{\Omega_{\rm D} t}) \mathbf{e}_Y \right] \; , \label{7.13}
\end{equation}
or
\begin{equation}
  X(t) = X(0) + \frac{V_0}{\Omega_{\rm D}} \sin{\Omega_{\rm D} t} \; , \label{7.14}
\end{equation}
\begin{equation}
  Y(t) = Y(0) + \frac{V_0}{\Omega_{\rm D}} (1 - \cos{\Omega_{\rm D} t}) \; , \label{7.15}
\end{equation}
where $\mathbf{R}(0)$ is aperture position. Magnetic deflection (\ref{7.15}) is determined by condition
\begin{equation}
  X(t) - X(0) = \frac{V_0}{\Omega_{\rm D}} \sin{\Omega_{\rm D} t} = 0 \; , \label{7.16}
\end{equation}
whence $\Omega_{\rm D} t = \pi$. The equation for determining the initial velocity $V_0$, if the magnetic deflection is known, has the form
\begin{equation}
  Y(t) - Y(0) = \frac{2 V_0}{\Omega_{\rm D}} = 2 \rho_0 = - \frac{2 m_0 V_0 \gamma(V_0)}{B_Z} \; , \label{7.17}
\end{equation}
where $\rho_0$ is the radius of the circle, along which the electron moves. In particular, in the classical case $\gamma(V_0)=1$ and the velocity is
\begin{equation}
  V_0 = - \frac{\rho_0 B_Z}{m_0} \; , \label{7.18}
\end{equation}
and for the relativistic case, where $\gamma(V_0) = (1-\beta^2_0)^{-1/2}$, we have
\begin{equation}
  V_0 = \frac{-\rho_0 B_Z}{m_0 \sqrt{1 + \cfrac{\rho^2_0 B^2_Z}{m^2_0 c^2}}} \; . \label{7.19}
\end{equation}
To obtain the standard formulas for magnetic deflection the component of magnetic vector $B_Z$ in eqs.~(\ref{7.18})-(\ref{7.19}) should be replaced by $eB_Z$, where $B_Z$ is magnetic field, and $e=-1$ (or $e = -1,6 \cdot 10^{-19}$ C in SI) is a helicity (or charge) of the electron. In general, the value of $V_0$ depends on the form of the function $u(V)$ according to eq.~(\ref{7.17}).

By the Kaufmann's method the initial velocity $V_0 = c \beta_0$ may be determined from equations for electric and magnetic deflections (\cite{Kau6}, S.~529, eqs.~(14)-(15)), which in our notations can be written as
\begin{equation}
  y' = L z(t) = \frac{e}{m_0} \frac{E}{c^2} \frac{1}{\beta^2_0 \gamma(\beta_0)} \; , \label{7.20}
\end{equation}
\begin{equation}
  z' = Y(t) - Y(0) = \frac{e}{m_0} \frac{M}{c} \frac{1}{\beta_0 \gamma(\beta_0)} \; , \label{7.21}
\end{equation}
where the electric field integral $E$ may be taken as $E=E_Z$, $z(t)$ is given by eqs.~(\ref{6.3}), (\ref{6.4}) or (\ref{6.5}), $e$ is the generally accepted absolute value of the electron charge. Comparison of eq.~(\ref{7.21}) with eq.~(7.17) shows that the magnetic field integral should be taken to be $M = 2 m^2_0 c^2 \beta^2_0 \gamma^2(\beta_0)/B_Z$, rather than $M=B_Z$, as it follows from Kaufmann's analysis. This means that $M$ is dependent on the initial velocity. Thus, it can be concluded that the expressions (\ref{6.4}) and (\ref{7.17}) for relativistic electron are incompatible with Kaufmann's formulas (\ref{7.20}) and (\ref{7.21}).

\section{Conclusion}

\renewcommand{\thesection}{\arabic{section}}

The results of this study, which cannot be considered totally exhaustive, show that the existence of spin greatly affects the trajectory of a charged particle even in stationary homogeneous fields. The incompleteness of the study is due not only with finding the solutions of some differential equations, but above all with the problem of determining the potential, which depends on the state of motion of spinning particle. Even in the simplest case, when the potential depends only on the relative distance and velocity, consideration of motion may be carried out without of using the ideas of special relativity, in which connection the speed limit to be possible in very special cases. However, input equations can be easily written in covariant form in the space with any number of spatial and temporal dimensions with any metric.

A special problem is the radiation of accelerated charged particles, which was not considered. But as follows from the energy balance, such radiation, apparently, is possible only in the presence of the relevant external fields.

Another problem is the equation of motion of spin that was used in the form (\ref{2.23}). It was shown in the case of two body problem, that it should be modified in the form $\mathbf{\dot{s}} = [\mathbf{\Omega}(t) \times \mathbf{s}] + \mathbf{m}(t)$, where the additive $\mathbf{m}(t)$ is due to external field, \cite{Tar4}. Therefore, it should be taken into account in Sec. 7 when considering the magnetic deflection. In this connection there has been suggested that the time to the asymptotic behavior is much more than the time to reach the deflection, allowing ignore $\mathbf{m}(t)$. Most likely, in strong fields this assumption will wrong.

Note also that the used equations of motion do not depend on the structure of  spinning particles, whose charge is determined by its helicity, while the structure should be manifested in the interaction with other particles.

\end{document}